\documentclass[journal]{IEEEtran}
\usepackage{cite}

\ifCLASSINFOpdf
\else
\fi

\usepackage{float}
\usepackage[utf8]{inputenc}
\usepackage[acronym]{glossaries}
\usepackage{subfig}
\usepackage{graphicx}
\usepackage{xcolor}
\usepackage{siunitx}
\DeclareSIUnit{\dBm}{dBm}
\DeclareSIUnit[]{\vrms}{V_{rms}}
\DeclareSIUnit[]{\arms}{A_{rms}}

\usepackage[ruled,vlined]{algorithm2e}
\usepackage{textcomp}
\usepackage{siunitx}
\usepackage{booktabs}
\usepackage{pgfplots}

\pgfplotsset{compat=1.18}
\begin{document}
%
\title{Electric Field Attenuation Techniques for Inductive Wireless Charging of Medical Implants}
%
%
%

\author{Samuelle~Boeckx $^{1}$,
        Pieterjan~Polfliet $^{2}$,
        Lieven~De~Strycker $^{1}$, and~Liesbet~Van~der~Perre $^{1}$
        \\
        \small{$^{1}$ \quad KU Leuven Ghent,  
Campus Rabot, 9000 Ghent, Belgium 
 \\
$^{2}$ Cochlear Limited, Cochlear Technology Centre Belgium, 2800 Mechelen, Belgium 
}
\thanks{This  research  was  funded by  Flanders  Innovation and  Entrepreneurship  (VLAIO), grant number HBC.2021.0797 and Cochlear Technology Centre.}
}

%
%

\markboth{Journal of \LaTeX\ Class Files,~Vol.~14, No.~8, August~2015}%
{Shell \MakeLowercase{\textit{et al.}}: Bare Demo of IEEEtran.cls for IEEE Journals}
%



\maketitle

\begin{abstract}
Inductive wireless charging of implantable medical devices necessitates careful control of magnetic and electric field emissions to meet strict safety regulations while delivering sufficient power. When designing a comfortable wireless charger that can operate over distances ranging to 10~cm or more, it is difficult not to exceed the most stringent electric field limit of 83~V/m set by the Canadian RSS-102 in December 2023 to avoid \newacronym{ns}{NS}{nerve stimulation}\acrfull{ns}. This paper investigates electric field attenuation techniques for mid-range wireless power transfer at 6.78~MHz. Using \newacronym{fea}{FEA}{finite element analysis}\acrfull{fea} like Ansys \textregistered{} HFSS \texttrademark{}, three mitigation strategies are evaluated; (1) a high-permittivity dielectric shielding layer to absorb and redistribute electric fields, (2) multiple resonant tuning capacitors distributed along the transmitter coil to lower the voltage swing and confine high E-field regions, and (3) alternative coil-array transmitter topologies to spatially localize more confined E-fields. The results show that each technique significantly reduces the E-field magnitude without substantially affecting the H-field magnitude. Shielding the transmit coil with a high-permittivity dielectric material attenuates the peak E-field from its initial 1416~V/m to 496~V/m, approximately a 65\% reduction. Distributing the tuning capacitance into sixteen smaller capacitors yields a drop from the 1416~V/m to 231~V/m, approximately a 84\% reduction. Both techniques preserve the required 8~A/m magnetic field. 
The third technique, a two-by-two coil array transmitter reduced the E-field from its 1416~V/m to 990~V/m (around 30\% reduction), though with a slight magnetic field redistribution. All three methods combined, the E-field was successfully attenuated to 82~V/m, just below the strictest 83~V/m Canadian RSS-102 exposure limit for \acrshort{ns}, without compromising power transfer efficiency. This research thus demonstrates a feasible approach and framework to safely extend the application of wireless charging for medical implants.
\end{abstract}


%
\IEEEpeerreviewmaketitle

\section{Introduction}
%
%
%
%

\IEEEPARstart{W}{ireless} Power Transfer \newacronym{wpt}{WPT}{wireless power transfer}(\acrshort{wpt}) systems are increasingly used across industries, from the seamless charging of smartphones and wearables to the promise of cable-free \acrfull{ev} charging and the life-enhancing potential of charging and powering implantable medical devices.
The convenience of not having bulky charging cables to charge an \acrshort{ev} is crucial in today's city architecture. For medical devices, like pacemakers and cochlear implants, the idea of not having to replace an implanted battery or not having the need for a visible external battery significantly enhances patient comfort. 
Despite its growing adoption, the widespread deployment of \acrshort{wpt} remains constrained by the fundamental challenge of efficient power transfer across larger distances. As the separation and/or misalignment between transmitter and receiver increases, maintaining sufficient \newacronym{pte}{PTE}{power transfer efficiency}\acrfull{pte} becomes more difficult, especially without exceeding safety thresholds for \newacronym{emf}{EMF}{electromagnetic field}\acrfull{emf} exposure. Regulatory bodies such as the \newacronym{icnirp}{ICNIRP}{International Commission on Non-Ionizing Radiation Protection}\acrfull{icnirp} have established strict limits on electric field strength for \acrfull{ns} and \newacronym{sar}{SAR}{specific absorption rate}\acrfull{sar} to protect human health. 
As a result, \acrshort{wpt} system engineers must navigate a complex trade-off between performance and regulatory compliance. The \acrshort{sar} can be measured in a compliance lab, \acrlong{ns} however, is not easy to measure, therefore the Canadian RSS-102 published an E-field reference level of \SI{83}{V/m} to make sure \acrshort{ns} does not occur. This is currently the strictest limit in the regulatory landscape and therefore used in this research.

This paper investigates different techniques to selectively attenuate an E-field emitted by an inductive charger without significantly affecting the H-field magnitude. It provides a framework to adhere to stringent E-field regulations. An example design in the application of \acrlong{wpt} for medical implants is proposed to illustrate the effect of the different techniques.
The following section provides a state-of-the-art overview of current \newacronym{ipt}{IPT}{Inductive Power Transfer} \acrfull{ipt} technologies across different domains, with a focus on compliance strategies. Section \ref{techniques} describes the different mitigation techniques, Section \ref{methods} explains the simulation model and Section \ref{assessment} quantifies the effect of the techniques on the E-field through simulations. Section \ref{discussion} discusses the results and section \ref{conclusion} concludes the paper.

\section{State of the Art}
The most widespread wireless charging applications today are consumer devices like charging pads for smartphones, wearables, and electric toothbrushes. These systems typically use the Qi \textregistered{} standard, developed by the \newacronym{wpc}{WPC}{Wireless Power Consortium}\acrfull{wpc} which is nowadays already widely adopted. It is the global protocol for \acrshort{ipt} in consumer electronics. First released in 2010, Qi has evolved through multiple versions and currently at 2.2.1 as of July 2025 \cite{QiHistory}. Qi systems operate at frequencies ranging from \SI{100}{\kilo \hertz} to \SI{205}{\kilo \hertz} with wavelengths from \SI{2.72}{\kilo \meter} to \SI{1.4}{\kilo \meter}. As a result, these devices operate in the near field, where the magnetic field is more dominant than the electric field. 
The Qi standard relies on close coupling and thus efficient power transfer over typically \SI{2}{} to \SI{10}{\milli \meter}. The power transfer is usually around \SI{5}{\watt}, but lately can go up to \SI{25}{\watt} for tablets and other larger devices \cite{Qi}. 
The electric field emitted by these charging systems is relatively weak and tightly confined compared to the magnetic field. The coils are also often shielded with ferrite sheets which serves a dual purpose. It guides the magnetic flux toward the receive coil which boosts the efficiency and it partly absorbs stray electric fields. 
Additionally, many \acrshort{wpt} systems include foreign object detection. These control circuits shut down the charger system if a metal object that could concentrate fields and heat up is detected, indirectly preventing unintended field exposure~\cite{FOD}.

The SAE J2954 standard \cite{SAE} is the Qi equivalent for \newacronym{ev}{EV}{electric vehicle} \acrlong{ev}s. Wireless charging for electric vehicles is one of the most power-intensive WPT applications, with typical systems transferring \SI{3.3}{\kilo \watt} to \SI{11}{\kilo \watt} across air gaps of \SI{10}{} to \SI{25}{\centi \meter}. The SAE J2584/2 standard is even targeting a power transfer of \SI{500}{\kilo \watt} for heavy-duty vehicles. The most common approach uses \newacronym{mrc}{MRC}{magnetic resonance coupling}\acrfull{mrc} at a frequency of \SI{85}{\kilo \hertz}, an \newacronym{ism}{ISM}{industrial, scientific and medical}\acrfull{ism} band specifically allocated for this application. \acrshort{ism} frequencies are often used, however, wireless charging of \acrlong{ev}s can use frequencies ranging from \SI{30}{\kilo \hertz} to \SI{30}{\mega \hertz}.
The way the transmit and receive coils are set up for this application (transmitter on or in the ground and receiver attached to the bottom of the vehicle) creates a strong oscillating magnetic field in the region between the coils.
Because of the high power, \acrshort{ev} chargers are carefully engineered to constrain the intense fields to the space underneath the car. The large coil assemblies are typically supported by ferrite tiles and often include aluminum shielding layers to confine those fields \cite{EVshield}. Hence, the magnetic field is largely vertical and stays between the pads, and any associated electric fields are mostly trapped near the coil surfaces. However, given the scale, human exposure must be considered in various scenarios: a person sitting inside the car while charging, a bystander near the vehicle, or even someone reaching under the vehicle while charging or a cat running underneath the car.
To comply with exposure limits, \acrshort{ev} \acrshort{wpt} systems employ robust field control measures. Many designs include, as mentioned, large ferrite blocks that redirect the magnetic flux while also acting as a barrier to electric fields. Another technique is to use shielding loops or plates. A metal screen for instance can be integrated to cover the coil edges, which shunts the electric field without significantly impeding the magnetic coupling. 
Some high-end systems use additional coils or clever coil current phasing to cancel out leakage fields. Recent research showed that by adjusting the phase difference between coil segments, E-field emissions can be halved without extra hardware, purely through circuit tuning \cite{Phase50}. Many \acrshort{ev}'s also include, like lots of consumer devices, foreign object detection.

Currently, there is no established standard for wireless charging of medical implants. There is a multitude of interacting factors making it a complex topic, such as location and depth of implantation, type of implant, and of course the fact that the transfer happens transcutaneously. Excessive E-fields can induce currents in tissue, causing localized heating (\acrshort{sar}) or unintended \acrfull{ns}. The power levels of medical implants are usually modest, often in the range of milliwatts, and the distances to cover are typically small as well, from around a few millimeters to \SI{5}{\centi \meter} \cite{power}. Although the general public exposure limits were not originally aimed at implanted devices, charging systems are evaluated for \acrshort{sar} compliance and field strength to ensure patient safety. For instance, wireless implant chargers are tested in tissue phantoms or simulations to verify that the peak \SI{1}{\gram} or \SI{10}{\gram} \acrshort{sar} in tissue stays under the 1.6 or 2 W/kg limit and that any temperature rise is within medical device safety margins. Manufacturers must also meet \newacronym{emc}{EMC}{electromagnetic compatibility} \acrfull{emc} standards so that the \acrshort{wpt} operation does not interfere with the device’s function or other equipment, and vice versa. Lastly, field limits must also be respected, for H-fields this is \SI{42}{\deci \bel \micro \ampere / \meter} following the harmonised standard EN300330 by \newacronym{etsi}{ETSI}{European Telecommunications Standards Institute}\acrfull{etsi} and for E-fields \SI{83}{V/m} following the most strict standard, the Canadian RSS-102 by \newacronym{ised}{ISED}{Innovation, Science and Economic Development Canada}\acrfull{ised} \cite{etsi} \cite{RSS}. This last standard was released in December 2023 to avoid \acrshort{ns}. Given that this effect is difficult to measure, the limit acts as a reference level.

Present-day implantable \acrshort{wpt} designs inherently minimize E-field emission by focusing on magnetic coupling. The transmit coils are often engineered to be electrically small and symmetric, reducing the voltage between coil elements and thus limiting E-fields. Many systems include a Faraday shielding or insulating layer between the coil and body to shunt or absorb electric fields without disturbing the magnetic flux. For instance, researchers have demonstrated that driving a pair of coils in an out-of-phase resonant mode can cancel much of the radiating E-field, significantly lowering the \acrshort{sar} \cite{OutOfPhase}. Likewise, biocompatible coatings on implants can serve to isolate the electric field from tissue, further cutting down localized \acrshort{sar} hot spots \cite{SARcoat}.
The challenge going forward is to maintain or improve power transfer efficiency for deeper implants without increasing E-field exposure.

\begin{figure}[!h]
\centerline{\includegraphics[width=\linewidth]{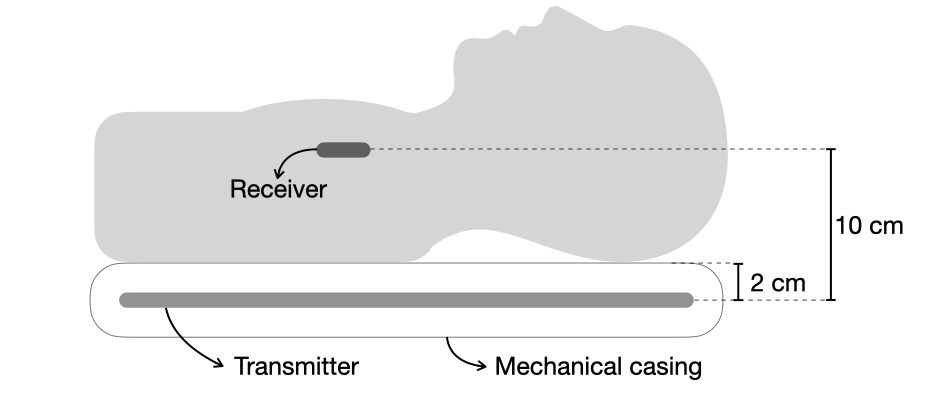}}
\caption{Situation sketch.}
\label{fig:situation}
\end{figure}

The goal of this research is to develop a comfortable charger for medical implants. As mentioned, these implants do not require a lot of power, as a benchmark we target \SI{100}{mW}. To charge comfortably, for instance when sitting or while sleeping, or to be able to charge implants that are implanted deep into the body, the achievable distance needs to be larger, up to \SI{10}{cm}. This is the distance at which we will measure the magnetic field magnitude throughout of the paper as a performance measure.
At this distance, coupling is weak so \acrfull{mrc} will be used in the design exploration. For the frequency of the magnetic field, we choose \SI{6.78}{MHz}, an \acrshort{ism} frequency. The electric field will be measured at a distance of \SI{2}{cm}, as this is the closest that human tissue can get to the transmit coils, because of the casing of the charger. A situation sketch is shown in Fig. \ref{fig:situation}.
Previous tests and measurements of potential designs of the charger system showed that the E-field limit of \SI{83}{V/m} is greatly exceeded. This paper will examine different strategies to attenuate this E-field while preserving the \acrlong{pte}.

\section{Mitigation Techniques} \label{techniques}
There are some known techniques to reduce electric fields in applications like inductive \acrlong{wpt} reported in literature. This section elaborates on the most promising and feasible techniques.

\subsection{Shielding} \label{sec:shield}
Shielding might be the most straight-forward technique when trying to reduce a field. The challenge, however, is that only the electric field should be shielded and not the magnetic field.
When looking for a material to shield with, there are certain properties to take into account like the dielectric loss tangent ($\tan \delta_e$), the magnetic loss tangent ($\tan \delta_m$), the relative permittivity $\epsilon_r$ and relative permeability $\mu_r$.

The dielectric loss tangent ($\tan \delta_e$) is the ratio of a material’s resistive (lossy) permittivity component to its reactive (energy-storing) component (tan $\delta_e = \frac{\epsilon\prime\prime}{\epsilon\prime}$). This parameter quantifies how strongly an electric field is attenuated within the material. A small $\tan \delta_e$ indicates that the material dissipates very little energy as heat, so an alternating electric field can largely oscillate without dying out. A high $\tan \delta_e$ means that the material dissipates a significant portion of the energy as heat. The material heats up and an \newacronym{ac}{AC}{alternating current}\acrfull{ac} electric field is strongly attenuated as it passes through. In capacitors or high-frequency devices, a large dielectric loss tangent lowers the \newacronym{q}{Q}{quality factor}\acrfull{q} and efficiency by damping the stored electric field energy \cite{dielectric}.

The magnetic loss tangent ($\tan \delta_m$), analogous to the dielectric loss tangent, is defined as the ratio of the imaginary (loss) part of permeability to the real part ($\tan \delta_m = \frac{\mu\prime\prime}{\mu\prime}$). It characterizes magnetic energy dissipation in the material, for instance due to hysteresis or eddy currents in a ferrite core. A low $\tan \delta_m$ ($\approx 0.0001$) means the material is magnetically nearly lossless, so an alternating magnetic field can pass through with minimal energy converted to heat, which is important for high-Q inductors. A high $\tan \delta_m$ ($\geq 0.1 $) indicates that the material significantly damps a changing magnetic field. The magnetic flux lags and collapses due to losses, manifesting as heat, for instance a ferrite with high loss tangent will heat up under \acrshort{ac} excitation and reduce an inductor’s \acrlong{q} \cite{mlosstan}.

The relative permittivity $\epsilon_r$ or dielectric constant is defined as the ratio of its permittivity to that of vacuum. 
The value of $\epsilon_r$ of vacuum is one, of other materials it is usually larger than one. It describes how a material responds to an electric field. A higher $\epsilon_r$ means that the material polarizes more in an electric field, storing more electric energy 
for a given charge, compared to air or vacuum. In practical terms, a high $\epsilon_r$ concentrates the electric field lines in the material, increasing the capacitance.
For instance, the dielectric to make a capacitor is always a high $\epsilon_r$ ($\approx 100$) material \cite{dielectric}.

The relative permeability ($\mu_r$) is the magnetic counterpart to $\epsilon_r$, defined as the ratio of a material’s permeability to the permeability of free space ($\mu_0$). It quantifies how easily the material supports the formation of magnetic fields within itself. A high $\mu_r$ material, like ferrite for instance, greatly concentrates magnetic flux lines, increasing inductance and magnetic energy storage for a given magnetizing force. In other words, the material concentrates the magnetic field compared to what it would be in air, which is why high-$\mu_r$ cores are used in inductors and transformers. A $\mu_r$ near 1 means the material has little effect on magnetic fields, like non-magnetic or air, while $\mu_r<1$ slightly repels magnetic fields \cite{perm}.

To conclude, relative permittivity and permeability concentrate the electric and magnetic field respectively, by storing field energy in the material, whereas loss tangents attenuate those fields by dissipating energy. So, to weaken an E-field without altering an H-field, a layer of material with a high $\epsilon_r$ and $\tan \delta_e$ and a low $\mu_r$ and $\tan \delta_m$ is needed in between the transmit coil and human tissue.

\subsection{Distributed Tuning Capacitors} \label{sec:C}
\begin{figure}[h!]
\centerline{\includegraphics[width=\linewidth]{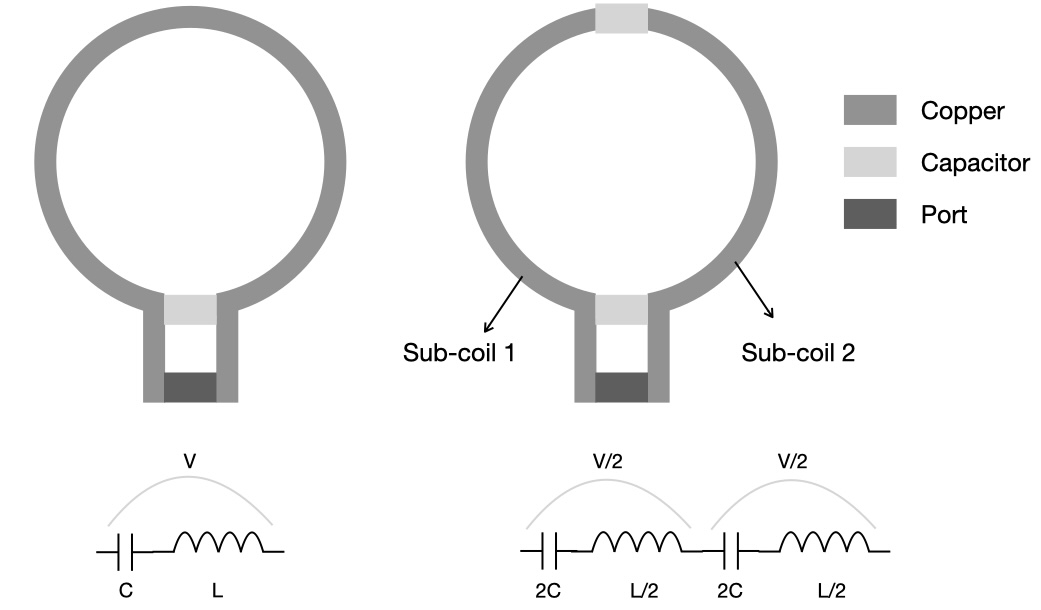}}
\caption{Configuration and circuit model of an example of the distributed capacitor technique.}
\label{fig:dis}
\end{figure}

One of the causes of the high electric field is the fact that the system is a series tuned resonant circuit. Due to the resonance, the voltage between the feed points of the coil reaches up to \SI{400}{\vrms}. This high voltage directly leads to strong electric fields in the vicinity of the coil.
A potential mitigation strategy is to split the tuning capacitor into multiple series capacitors that are distributed along the coil turns, thereby forming multiple sub-coils. Such an approach was proposed by
A. Nezaratizadeh et al. to reduce the \acrshort{sar} during wireless charging of implantable medical devices \cite{Cap}. Since \acrshort{sar} is primarily an effect of an electric field, their results indicate that this distributed capacitors approach can effectively reduce electric fields.
An important aspect of this method is that, if each sub-coil has an identical inductance, the corresponding tuning capacitors are also identical. As a result, each resonant sub-tank exhibits the same voltage swing, equal to a fraction of the total resonant voltage. This voltage division reduces the peak electric field generated by every resonant sub-tank significantly compared to a single resonant tank. Additionally, distributing the tuning capacitors geometrically far away from each other increases the physical separation between high-voltage nodes, which further reduces local electric field concentration between adjacent capacitors. A visualization of this concept is shown in Fig. \ref{fig:dis} 

\subsection{Different Coil Topologies} \label{sec:coiltop}
Another cause of high electric fields in mid‑range wireless chargers is simply the physical size of the transmit coil. Large coils cover a wide area 
which increases the extent of the surrounding electric field. Smaller coils, by contrast, confine the electric field more tightly. One design approach is to construct the charging pad as a grid or array of smaller coils, each covering only a small portion of the total area and operating at a lower voltage 
During operation, only the coils located directly beneath or near the implanted device can be energized, thereby concentrating the magnetic field where it is needed while keeping the rest of the pad essentially field‑free. This selective activation method not only minimizes unnecessary electric field exposure but can also improve overall system efficiency by avoiding power dissipation in unused regions of the pad.



\section{Simulation Model} \label{methods}
A simulation-based assessment is conducted to quantify the effect of these mitigation techniques on the electric field.
A simplified model of a wireless charging coil is created using Ansys HFSS \cite{ansys}, a \acrfull{fea} simulation tool. In order to make a resonant system at \SI{6.78}{MHz}, the self-inductance of the coils should not be too large as this results in smaller capacitor values, which makes the system sensitive to parasitic capacitance. Inductances in the order of \SI{1}{} to \SI{10}{\micro \henry} result in capacitor values in the range of \SI{550}{} to \SI{55}{\pico \farad}, which is feasible.

\begin{figure}[h!]
\centerline{\includegraphics[width=\linewidth]{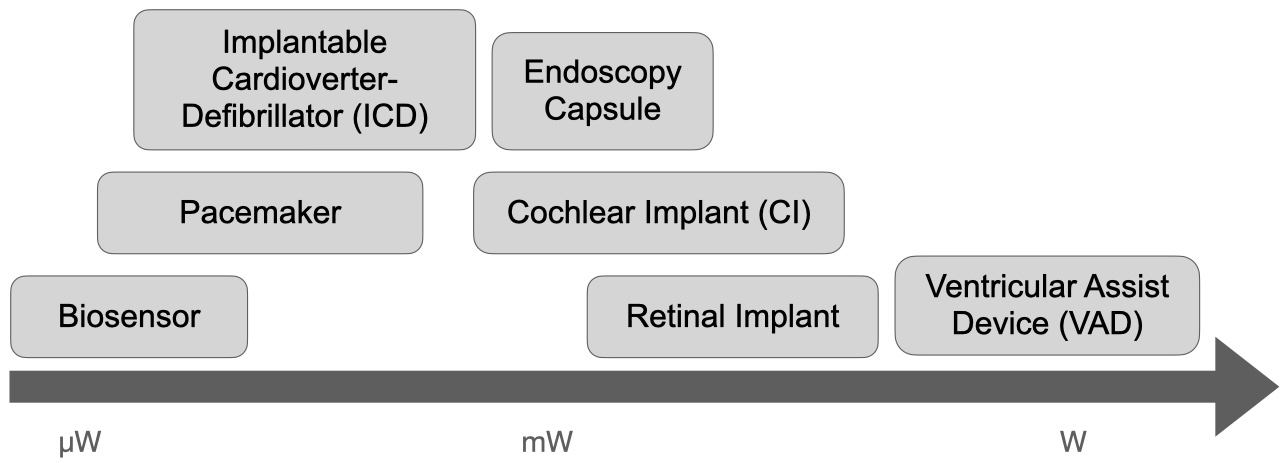}}
\caption{Range of power requirements of implantable medical devices \cite{b3}.}
\label{fig:power}
\end{figure}

Looking at Fig. \ref{fig:power}, most types of active medical implants can be powered if the transmitter can deliver up to a few hundreds of milliwatts.
Most implants have similar dimensions, with diameters of around \SI{3}{\centi \meter}. A transmitter that creates a charging region of around \SI{50}{cm} by \SI{25}{cm} should be more than sufficient to comfortably cover abdomen, chest or head, where these implants from Fig.\ref{fig:power} are usually located. The initial idea is to use two coils to cover this area to avoid a too high inductance and to make sure the magnetic field is homogeneous enough. There is a small overlap between the coils, as this alignment ensures minimal coupling between the two transmit coils.
We use this geometry of the wireless charger throughout the remainder of the paper.

We further assume, as mentioned, a worst-case vertical distance of \SI{10}{cm} between transmit and receive coil and no angular misalignment i.e. the coils are in parallel .
The \acrfull{pte} $\eta$ can be calculated using \eqref{link} \cite{WPT}:
\begin{equation}
    \eta = \frac{P_R}{P_T + P_R} = \frac{k^2 Q_T Q_R}{1 + k^2 Q_T Q_R}
    \label{link}
\end{equation}
with $P_T$ and $P_R$ the power of the transmit and receive coil,  $Q_T$ and $Q_R$ the quality factor of the transmit and receive coil and $k$ the coupling coefficient between the two coils. $Q_T$ was measured with an LCR meter, once the prototype for verification was built, and resulted in a value of 35. For $Q_R$, a smaller coil, 20 is a typical value. The coupling coefficient for a resonant coupled system is typically rather low, $\leq$ \SI{0.1}{} \cite{WPT}. Using the approach to estimate coupling coefficients, as presented in \cite{UAV}, $k$ is estimated to be around \SI{0.005}{}. This results in a link efficiency of \SI{1.7}{\%} and thus a required transmit coil power of \SI{5.9}{\watt}. If a \SI{5}{V} driver is used, this results in a current of \SI{1.2}{\arms} through the transmit coil. This current will flow through all simulated transmit coils.

\begin{figure}[H]
\centerline{\includegraphics[width=\linewidth]{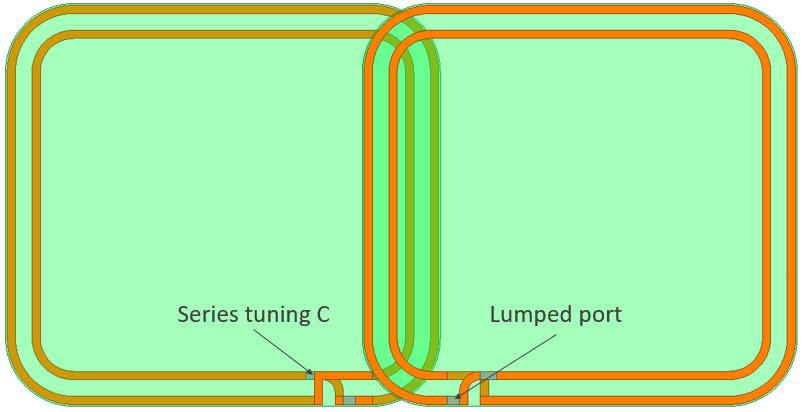}}
\caption{Simulated structure of the charger prototype.}
\label{fig:view}
\end{figure}

\begin{figure}[H]
\centerline{\includegraphics[width=\linewidth]{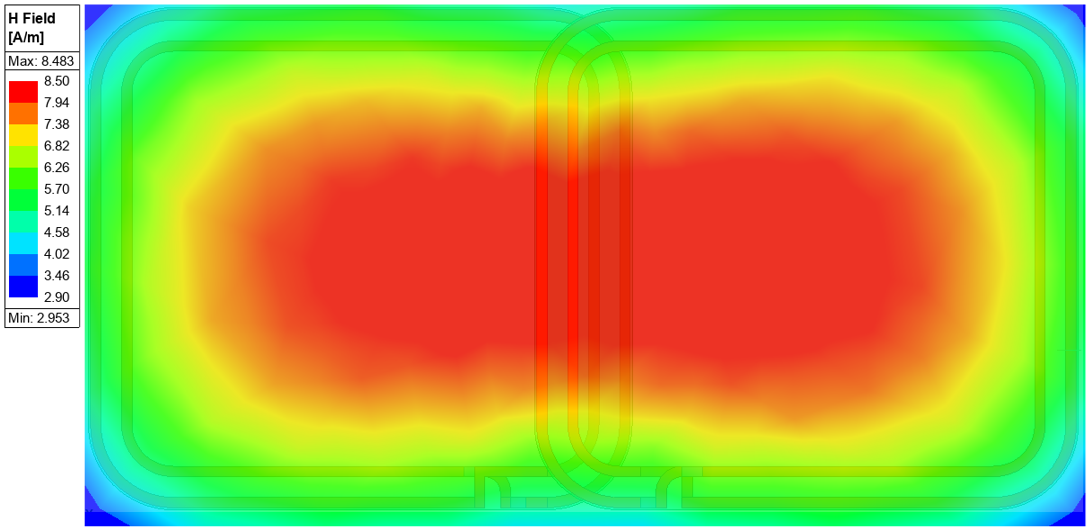}}
\caption{Simulations of the magnetic field of the charger prototype at a height of 10 cm.}
\label{fig:H}
\end{figure}

\begin{figure}[H]
\centerline{\includegraphics[width=\linewidth]{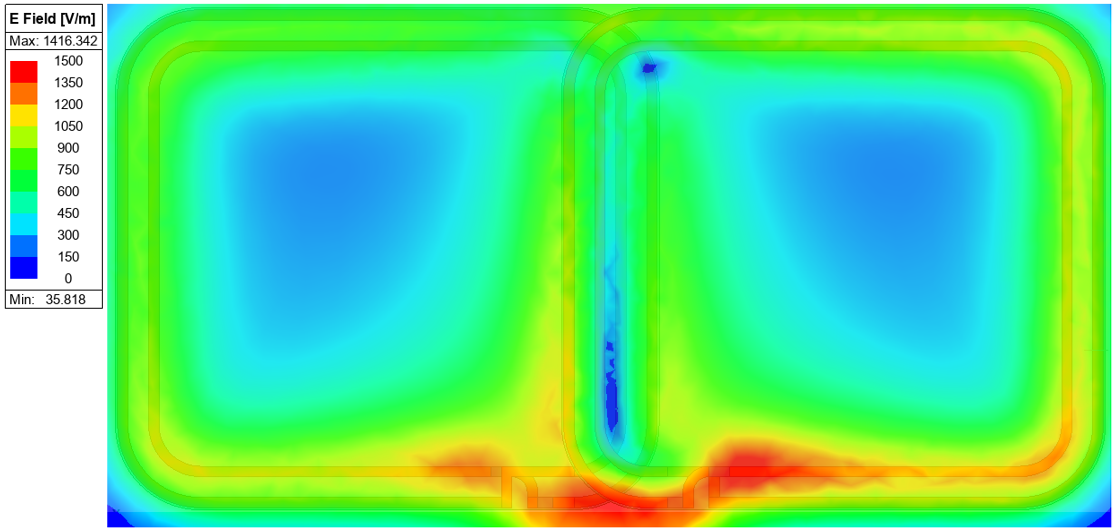}}
\caption{Simulations of the electric field of the charger prototype at a height of 2 cm.}
\label{fig:E}
\end{figure}

We calculate the voltage induced over the receive coil in a similar way. We assume five closely wound windings for the \SI{3}{\centi \meter} diameter receive coil. Ansys HFSS simulations result in a self-inductance of \SI{1.23}{\micro \henry}. The quality factor is, as previously mentioned, 20, so, this results in an equivalent resistance of \SI{2.62}{\ohm}. Taking into account the low coupling factor, the receive coil can be represented by a voltage source (i.e. the induced emf) in series with a source resistance of \SI{2.62}{\ohm} \cite{UAV}. In order to have a maximum power transfer to the load of the receive coil (i.e. the implant), the load resistance should be equal to the source resistance. Delivering a power of \SI{0.1}{\watt} to the implant requires a voltage of \SI{0.512}{\volt} over this load. So, the emf described in Faraday's law should be \SI{1.02}{\vrms}.
A changing magnetic field through a surface induces an emf in any boundary path of that surface \cite{Fleisch}. The relation between this electromotive force and the magnetic flux $\phi$ is given by Faraday's law \eqref{eq:emf}:
\begin{equation}    
    emf = N \frac{d\phi}{dt}
    \label{eq:emf}
\end{equation}
with $N$ the number of windings of the receive coil. The magnetic flux $\phi$ can be defined as the part of the magnetic field that is crossing a surface \cite{Fleisch}, the receive coil in this case. It is defined by Gauss' law of induction in equation \eqref{eq:flux}:
\begin{equation}
\phi = \int \Vec{B} \cdot d\Vec{a}
    \label{eq:flux}
\end{equation}
with $B$ the magnetic flux density ($B = \mu_0 H$) through surface $a$. 
We solve these equations for $H$, with the information assumed and calculated above, giving that the minimal magnetic field intensity needed to deliver \SI{100}{\milli \watt} to the implant is \SI{7.65}{\ampere / \meter}.
The transmit coils, seen on Fig. \ref{fig:view}, simulated in Ansys HFSS deliver a magnetic field like this, the maximum field magnitude at \SI{10}{cm} above the transmit coils reaches up to \SI{8.5}{A/m} as shown on Fig. \ref{fig:H}. The current through the coils with an applied voltage of \SI{10}{V} is simulated to be \SI{1.24}{\arms}. Unfortunately, the electric field that is also emitted by this transmitter has a value of \SI{1416}{V/m} at a height of \SI{2}{cm} as shown in Fig. \ref{fig:E}, around 17 times higher than what the Canadian standard RSS-102 allows. As a verification for these results, the electric field of a prototype that is represented by the simulations was measured with an electric field probe and spectrum analyzer. The maximum field strength was measured to be \SI{1381}{V/m}. We can thus conclude that the simulations will offer valuable insights on the effect of the mitigation techniques on the electric field strength.

\section{Assessment of Mitigation Techniques} \label{assessment}
The three techniques described in Section \ref{techniques} will be evaluated in Ansys HFSS to assess the effectiveness of attenuating this E-field without altering the H-field.

\subsection{Shielding}
\begin{figure}[H]
\begin{tikzpicture} 
\begin{axis}[
    xlabel={$\epsilon_r$},
    ylabel={Max E-field [V/m]},
    xmin=0, xmax=1000,
    ymin=450, ymax=1500,
    xtick={250,500,750,1000},
    ytick={500,750,1000,1250,1500},
    legend pos=north west,
    ymajorgrids=true,
    grid style=dashed,
]
\addplot[
    color=blue,
    mark=dot,
    ]
    coordinates {
    (0,1416)(25,1170)(50,1043)(75,969)(100,907)(200,757)(300,676)(400,598)(500,585)(600,582)(700,547)(800,556)(900,496)(1000,547)
    };
    \addlegendentry{$\tan \delta_e=0.01$}
\addplot[
    color=red,
    mark=dot,
    ]
    coordinates {
    (0,1416)(25,1161)(50,1029)(75,933)(100,887)(200,746)(300,669)(400,597)(500,585)(600,582)(700,552)(800,543)(900,505)(1000,546)
    };
    \addlegendentry{$\tan \delta_e=0.02$}
\addplot[
    color=orange,
    mark=dot,
    ]
    coordinates {
    (0,1416)(25,1167)(50,1048)(75,956)(100,887)(200,746)(300,670)(400,597)(500,585)(600,580)(700,551)(800,544)(900,517)(1000,545)
    };
    \addlegendentry{$\tan \delta_e=0.03$}
\addplot[
    color=green,
    mark=dot,
    ]
    coordinates {
    (0,1416)(25,1167)(50,1048)(75,956)(100,886)(200,745)(300,669)(400,597)(500,587)(600,579)(700,551)(800,543)(900,506)(1000,553)
    };
    \addlegendentry{$\tan \delta_e=0.04$}
\addplot[
    color=purple,
    mark=dot,
    ]
    coordinates {
    (0,1416)(25,1167)(50,1028)(75,955)(100,901)(200,745)(300,668)(400,596)(500,587)(600,579)(700,554)(800,543)(900,518)(1000,552)
    };
    \addlegendentry{$\tan \delta_e=0.05$} 
\end{axis}
\end{tikzpicture}
\caption{Maximum E-field value in function of the material's permittivity}
\label{gr:perm}
\end{figure}
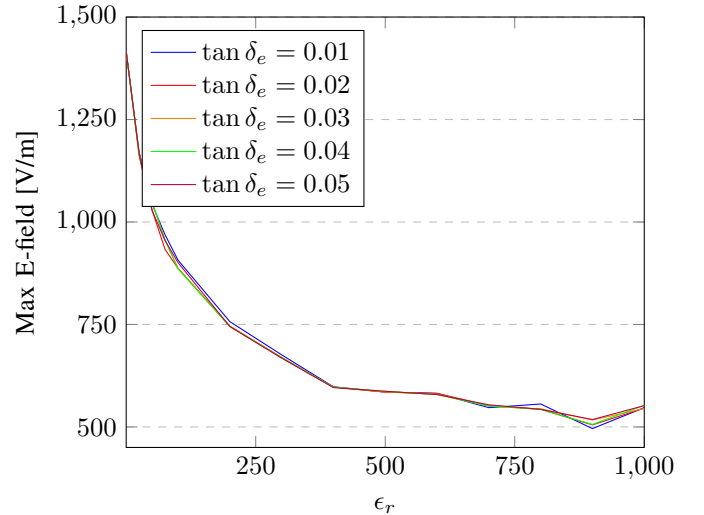

The parameter requirements in Section \ref{sec:shield} oblige us to look for a material with a high $\epsilon_r$ and $\tan \delta_e$ and a low $\mu_r$ and $\tan \delta_m$. 
When looking into different types of materials, we find that typical materials like that are water and saline. Water has an $\epsilon_r$ of \SI{80}{} around room temperature and a $\tan \delta_e$ ranging from \SI{0.01}{} to \SI{0.05}{} depending on the purity. For saline this is around \SI{70}{} and \SI{0.03} respectively. Values for these parameters, reported in literature range up to \SI{1000}{} for $\epsilon_r$ and around \SI{0.5}{} at \SI{6.78}{MHz} for $\tan \delta_e$.
To be able to search more effectively for the best material, it is beneficial to know which parameter has the greatest effect on the E-field. For this purpose, the E-field of the charger prototype was simulated with a layer of arbitrary material of \SI{480}{mm} by \SI{250}{mm}, this covers the prototype perfectly, and with a thickness of \SI{1}{mm} at a height of \SI{10}{mm} above the charging coils. We assumed for $\mu_r$ a value of 1 and for $\tan \delta_m$ \SI{0.0001}{}. The $\epsilon_r$ was varied from 1 till 1000 and $\tan \delta_e$ from \SI{0.01}{} till \SI{0.5}{}. The resulting maximum E-field values above the layer of material, corresponding to the right $\epsilon_r$, are shown in in Fig. \ref{gr:perm}.

\begin{figure}[H]
\centerline{\includegraphics[width=\linewidth]{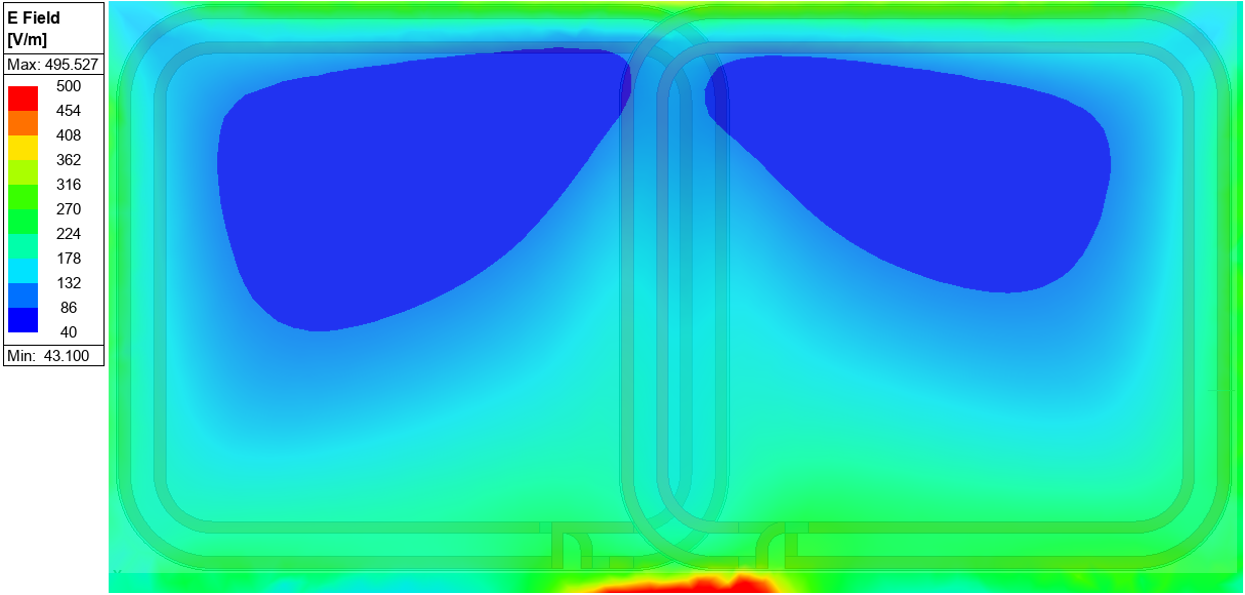}}
\caption{E-field of the charger with a \SI{1}{mm} layer of $BaTiO_3$.}
\label{fig:Eshield}
\end{figure}

From this graph it is clear that the $\epsilon_r$ has the largest effect on the E-field, we thus look for materials with a $\epsilon_r \approx 900$. 
Barium Titanate ($BaTiO_3$) fits the requirements best. 
Figure. \ref{fig:Eshield} shows the electric field with a layer of $BaTiO_3$ above the coils, it is reduced to 496 V/m. Although the E-field is significantly attenuated, it is not yet compliant to the RSS-102 standard. The next subsection studies the second technique to further reduce the E-field.

\subsection{Distributed Capacitors}
\begin{figure}[H]
\centerline{\includegraphics[width=\linewidth]{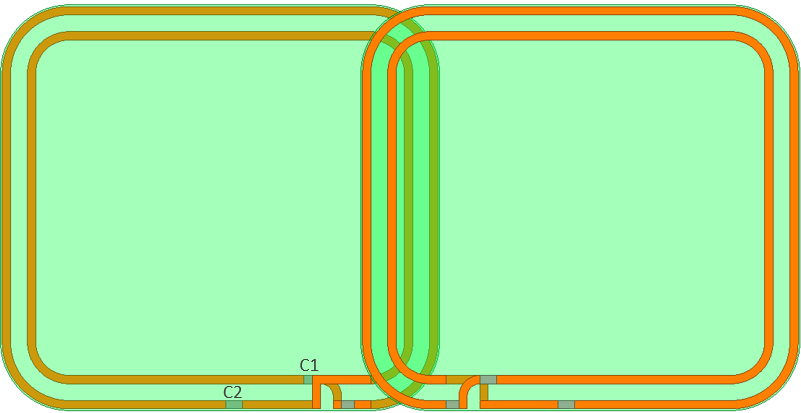}}
\caption{Simulated structure of the charger prototype with two capacitors.}
\label{fig:2C}
\end{figure}

\begin{figure}[H]
\centerline{\includegraphics[width=\linewidth]{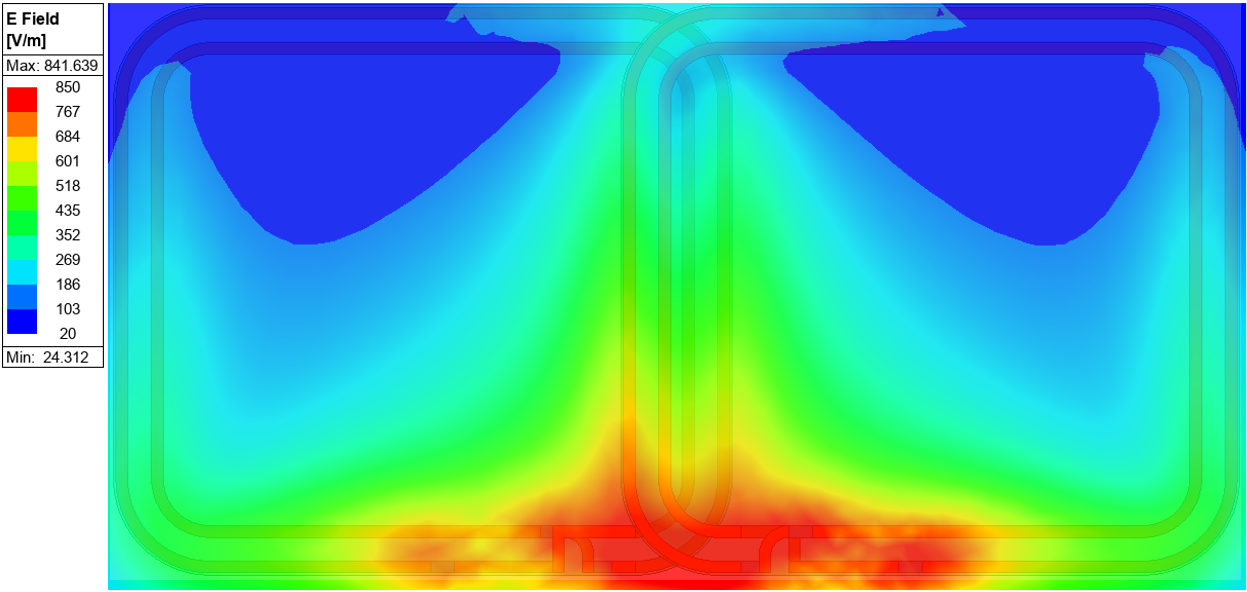}}
\caption{Simulations of the electric field of the charger prototype with two capacitors.}
\label{fig:E2C}
\end{figure}

\begin{figure}[H]
\centerline{\includegraphics[width=\linewidth]{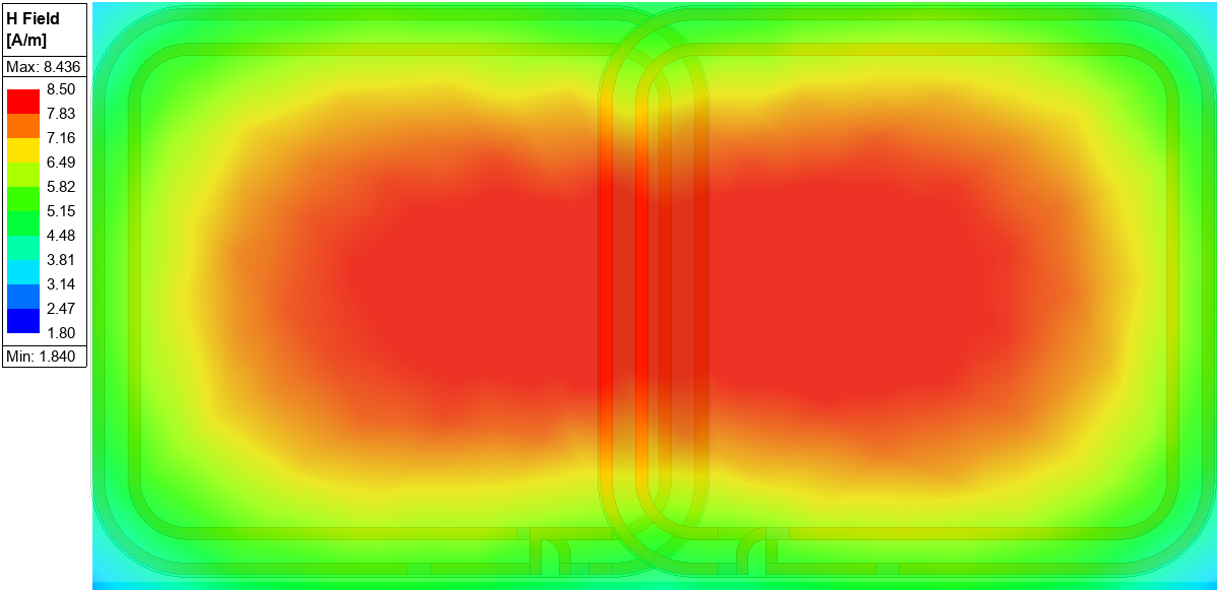}}
\caption{Simulations of the magnetic field of the charger prototype with two capacitors.}
\label{fig:H2C}
\end{figure}

When looking at Fig. \ref{fig:E}, the maximum E-field values are clearly located around the capacitors. This gives great confidence that the technique of distributing the capacitors along the coil traces, as described in Section \ref{sec:C}, can effectively lower the E-field.
To carry out this technique, the coils are divided in to two sub-coils by adding a capacitor. This results in a capacitor at the start of each turn of the coils, as represented by $C_1$ and $C_2$ shown in Fig. \ref{fig:2C}. We need to know the value of the inductance of each turn or sub-coil to be able to calculate the value of the tuning capacitors. We expect the inductance to be around half of the inductance of the full coil and the capacitor double the value of the initial tuning capacitor. However, there is also the mutual inductance between the turns that will certainly have an impact and seeing that the tuning of the system is crucial for the power transfer, this needs to be verified. Simulating the turns separately results in an inductance of \SI{0.74}{} $\mu H$ and \SI{0.86}{} $\mu H$ for the inner and outer turn respectively and \SI{0.38}{} $\mu H$ for the mutual inductance between the turns. To calculate the tuning capacitors for each turn, the following formula \eqref{eq:C} is used:

\begin{equation}
\label{eq:C}
    f = \frac{1}{2\pi\sqrt{(L+M)C}}
\end{equation}

with f the tuning frequency namely \SI{6.78}{MHz}, L the inductance of every turn, M the mutual inductance between the turns and C the tuning capacitor, $C_1$ or $C_2$ in this case. We add this value to the inductance of the turns, resulting in \SI{1.12}{} $\mu H$ and \SI{1.24}{} $\mu H$. We find for $C_1$ a value of \SI{490}{pF} and for $C_2$ \SI{444}{pF}. A spice AC analysis shows that the system is then indeed perfectly tuned to \SI{6.78}{MHz} again. 
When we put these values into Ansys HFSS again, we see a clear reduction in the electric field magnitude. As seen on Fig.~\ref{fig:E2C}, the maximum field is now reduced to \SI{842}{V/m}, while the magnetic field remained unchanged, as seen in Fig. \ref{fig:H2C}. Seeing that the two capacitors are geometrically still located close together, the max E-field is still located  at the same place as on Fig. \ref{fig:E}, but already a bit more stretched out.

\begin{figure}[H]
\centerline{\includegraphics[width=\linewidth]{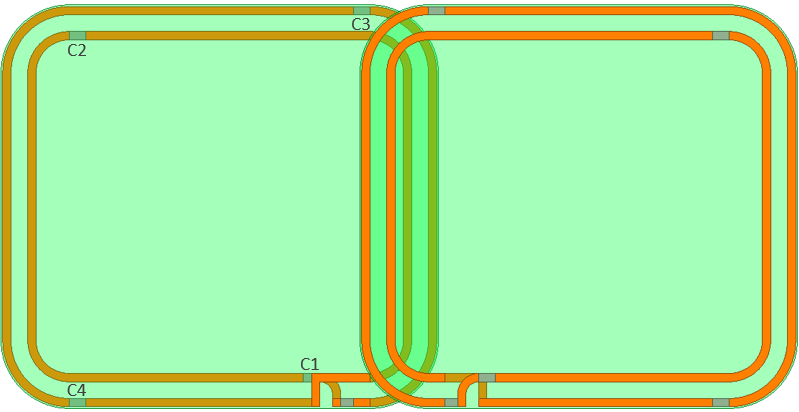}}
\caption{Simulated structure of the charger prototype with four capacitors.}
\label{fig:4C}
\end{figure}

This reduction of around 30\% compared to the initial prototype is quite significant and gives strong confidence that adding even more tuning capacitors could distribute the electric field even more. To put this to practice, we followed the same procedure as implementing the two distributed capacitors, but now with four capacitors per coil, one every half a turn. As seen on Fig. \ref{fig:4C} the capacitors aren't placed at exactly every half a turn as this would result in having two times two capacitors geometrically close together. Simulations showed that if this is the case, a rather high E-field can again be seen at this location. Spreading them out geometrically results in a higher reduction of the E-field than having the sub-coils and thus tuning capacitors at exactly the same inductance and capacitance value.
The capacitors C1, C2, C3 and C4 now have values of \SI{801}{pF}, \SI{872}{pF}, \SI{709}{pF} and \SI{806}{pF} respectively. The electric field was now further reduced to \SI{519}{V/m} while the magnetic field still reaches a maximum magnitude of around \SI{8}{A/m} and the shape did not noticeably change.

With eight capacitors, the electric field gets as low as \SI{376}{V/m} without altering the magnetic field. The tuning of the system gets a little more involving as there are 28 mutual inductances to take into account now. For this prototype we thus assume that the value of the tuning capacitors is around double that of the tuning capacitors of the prototype with four capacitors. After some experimenting, a value of \SI{2014}{pF} for all eight capacitors, resulted in perfect tuning at \SI{6.78}{MHz}.

\begin{figure}[H]
\centerline{\includegraphics[width=\linewidth]{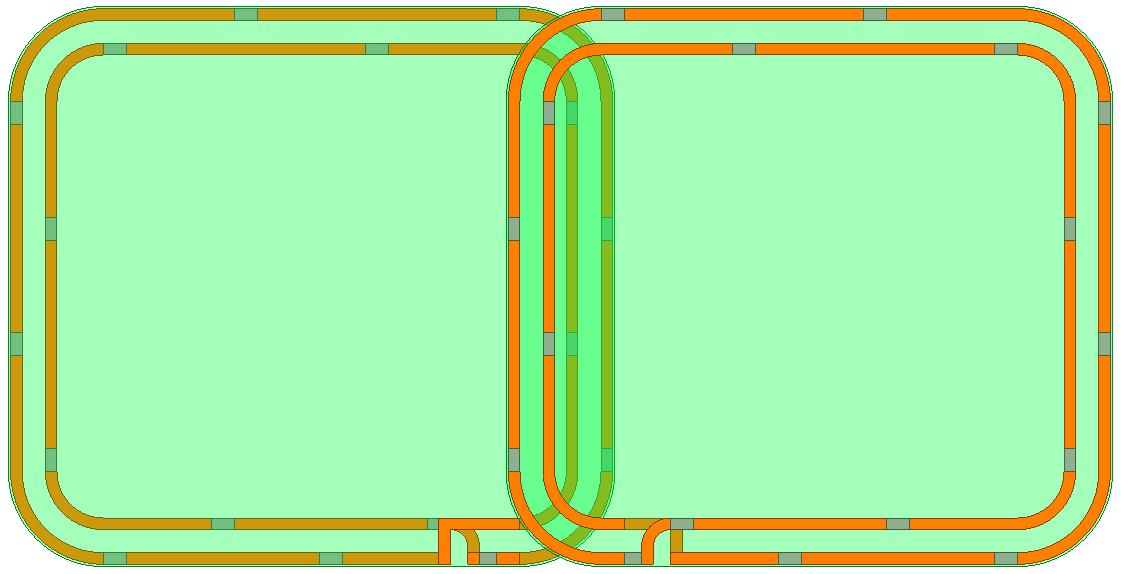}}
\caption{Simulated structure of the charger prototype with sixteen capacitors.}
\label{fig:moreC}
\end{figure}

\begin{figure}[H]
\centerline{\includegraphics[width=\linewidth]{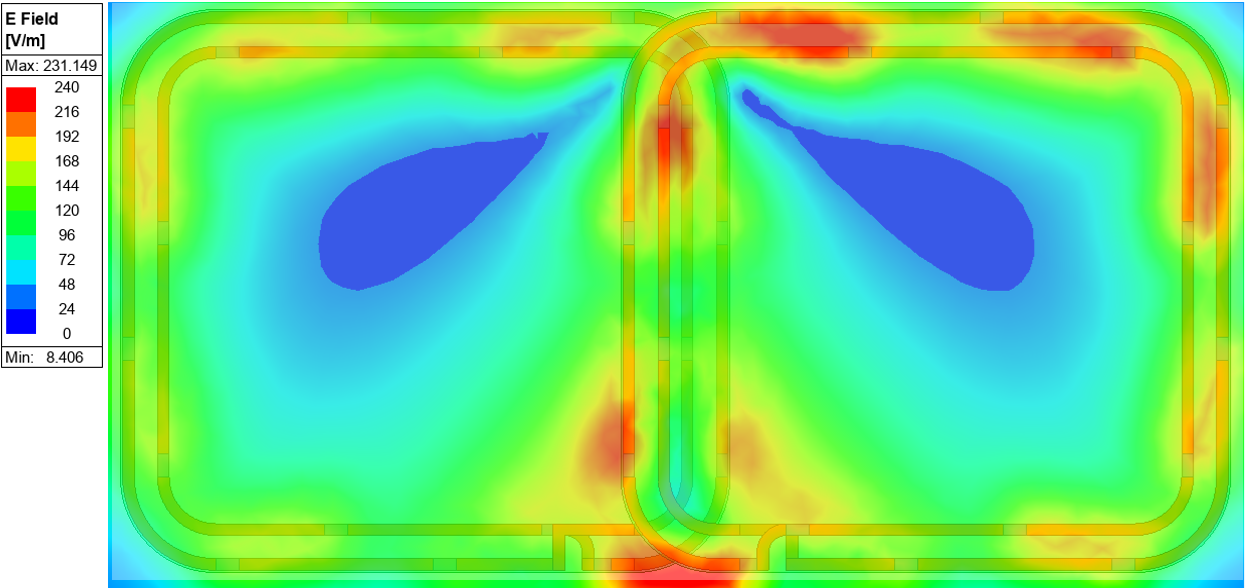}}
\caption{Simulations of the electric field of the charger prototype with sixteen capacitors.}
\label{fig:E16C}
\end{figure}

With sixteen capacitors, the electric field gets as low as \SI{231}{V/m} without altering the magnetic field. The tuning of the system is even more cumbersome, so we assume again around double the capacitance as the previous prototype with eight capacitors. A value of \SI{4100}{pF} tuned the system perfectly again. In Fig.~\ref{fig:moreC} and \ref{fig:E16C}, we can now very clearly see that the high electric field magnitude spots are exactly where capacitors are located. The magnitude seems higher for the right coil than for the left coil. This is because the coils overlap and as a result the right coil is a little closer to the plane where the E-field is measured.

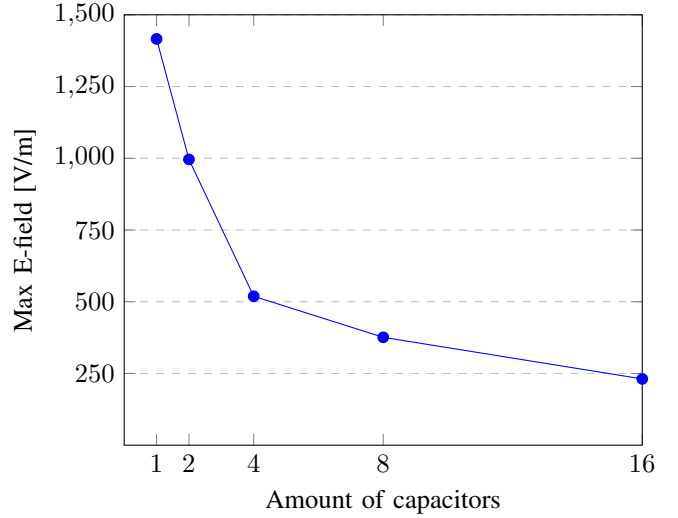
\begin{figure}[H]
\begin{tikzpicture}
\begin{axis}[
    xlabel={Amount of capacitors},
    ylabel={Max E-field [V/m]},
    xmin=0, xmax=16,
    ymin=0, ymax=1500,
    xtick={1,2,4,8,16},
    ytick={250,500,750,1000,1250,1500},
    legend pos=north west,
    ymajorgrids=true,
    grid style=dashed,
]
\addplot[
    color=blue,
    mark=*,
    ]
    coordinates {
    (1,1416)(2,996)(4,519)(8,376)(16,231)
    };   
\end{axis}
\end{tikzpicture}
\caption{Maximum E-field value in function of the amount of capacitors}
\label{gr:C} 
\end{figure}

If we plot the electric field magnitude in function of the amount of tuning capacitors like in Fig. \ref{gr:C}, we see that the plot resembles a 1/x pattern, we could thus keep adding capacitors but the reduction of E-field will become less and less while the effort of tuning gets more and more cumbersome. We therefor leave it at sixteen tuning capacitors and consider the third technique to reduce the electric field.

\subsection{Different Coil Topologies}
\begin{figure}[H]
\centerline{\includegraphics[width=\linewidth]{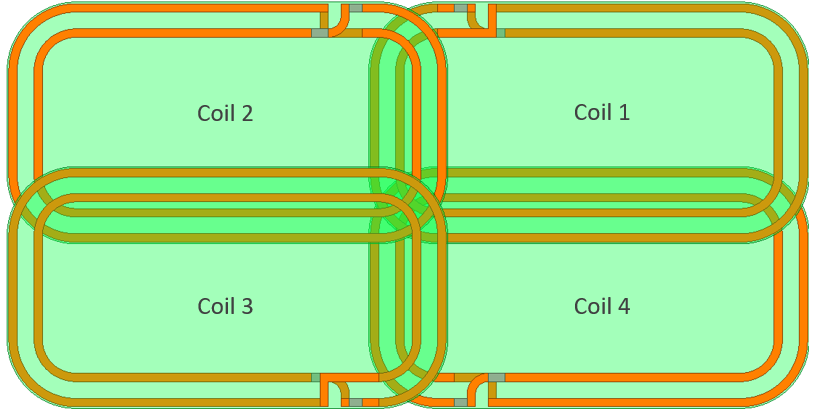}}
\caption{Simulation topology of the charger prototype with four coils.}
\label{fig:4coils}
\end{figure}

\begin{figure}[H]
\centerline{\includegraphics[width=\linewidth]{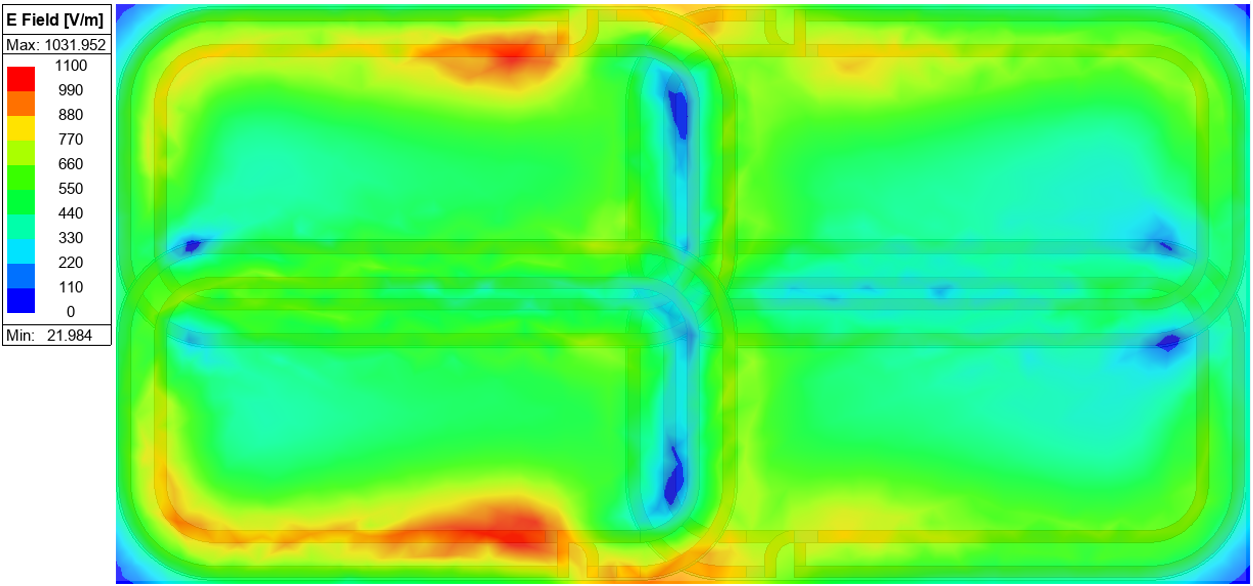}}
\caption{E-field of the charger prototype with four coils.}
\label{fig:E4coils}
\end{figure}

\begin{figure}[H]
\centerline{\includegraphics[width=\linewidth]{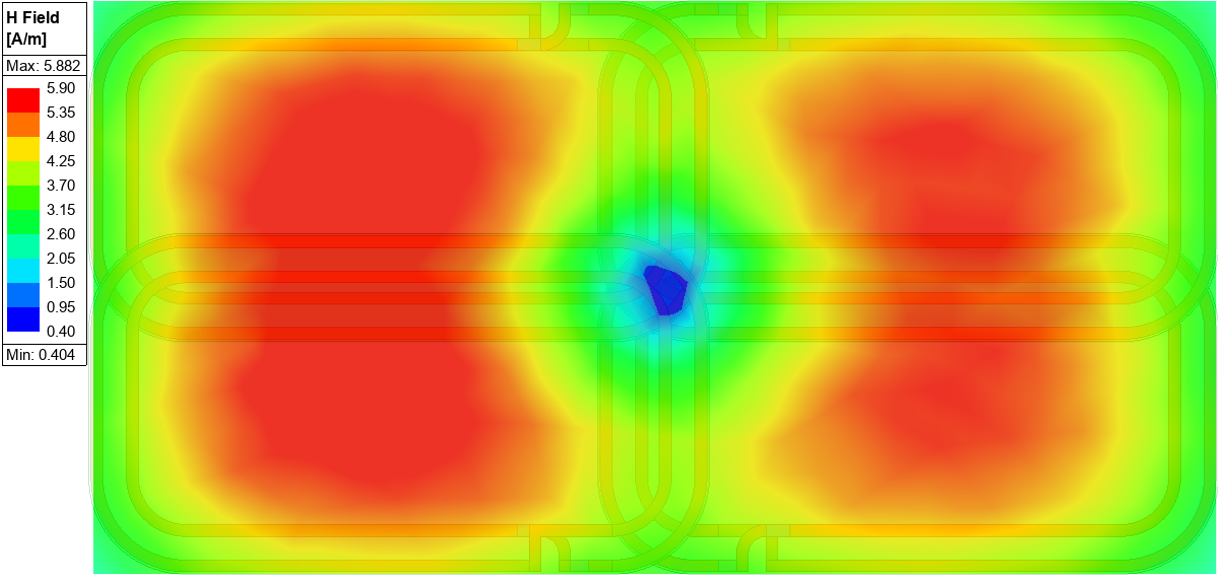}}
\caption{H-field of the charger prototype with four coils.}
\label{fig:H4coils}
\end{figure}

\begin{figure}[H]
\centerline{\includegraphics[width=\linewidth]{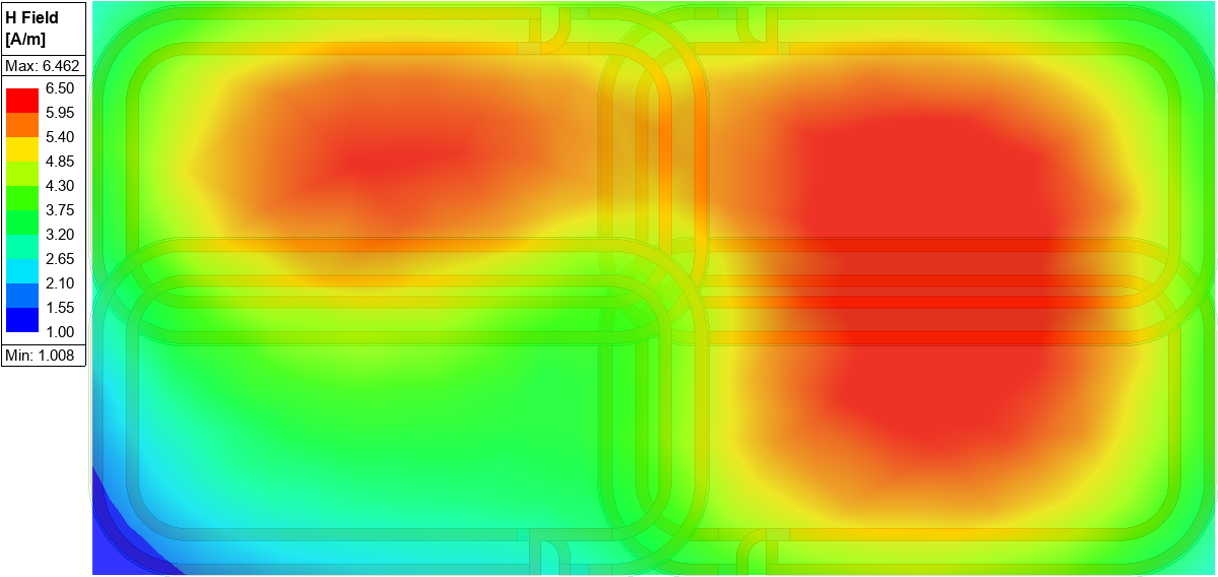}}
\caption{H-field of the charger prototype with coil 1 with a phase difference of 90°.}
\label{fig:phase1}
\end{figure}

\begin{figure}[H]
\centerline{\includegraphics[width=\linewidth]{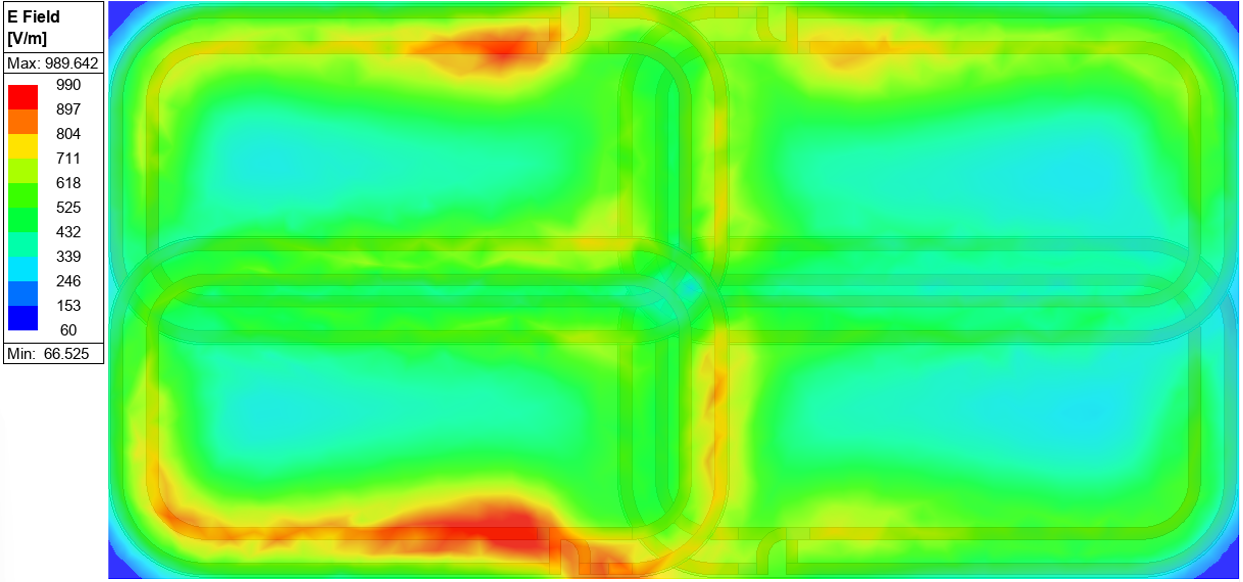}}
\caption{E-field of the charger prototype with coil 1 with a phase difference of 90°.}
\label{fig:phase1E}
\end{figure}

\begin{figure}[H]
\centerline{\includegraphics[width=\linewidth]{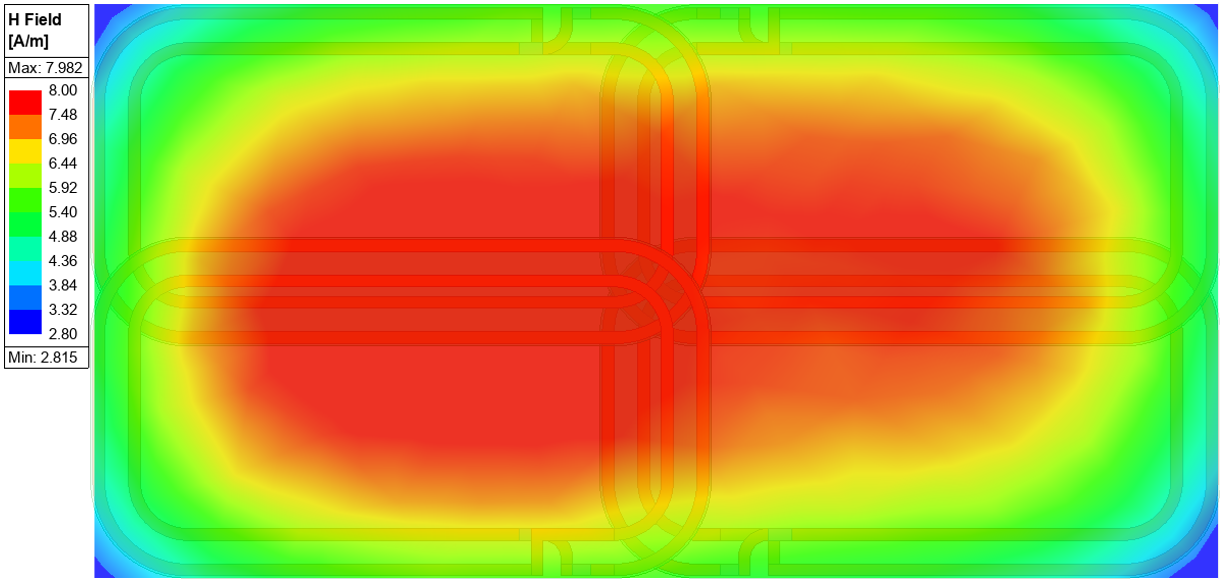}}
\caption{H-field of the charger prototype with coils 1 and 3 with a phase difference of 90°.}
\label{fig:phase12H}
\end{figure}

\begin{figure}[H]
\centerline{\includegraphics[width=\linewidth]{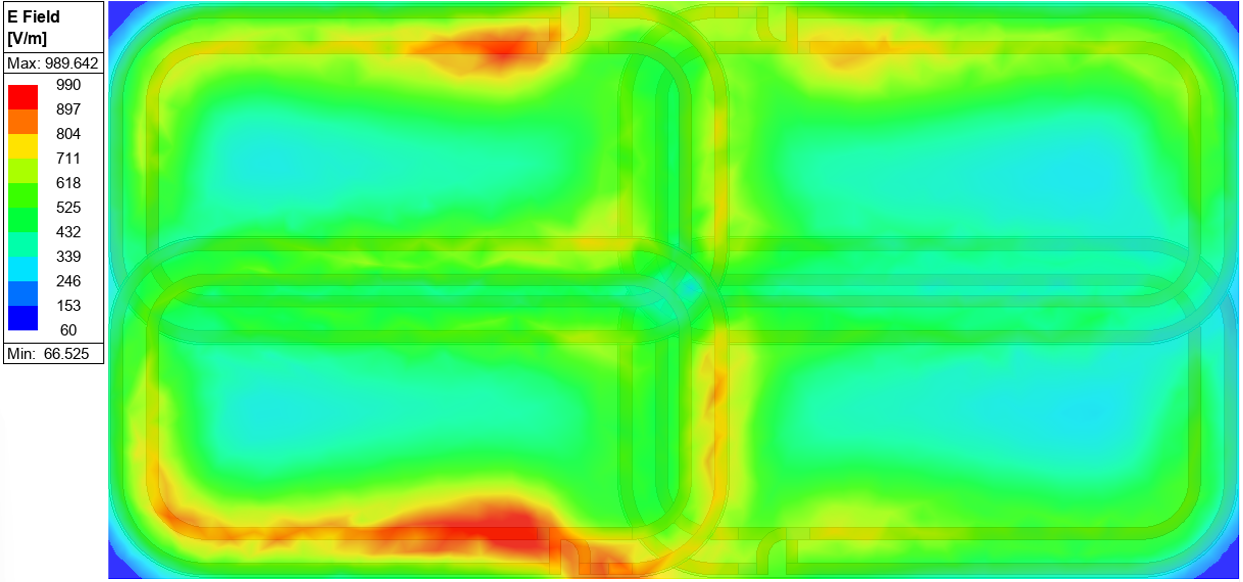}}
\caption{E-field of the charger prototype with coils 1 and 3 with a phase difference of 90°.}
\label{fig:phase12E}
\end{figure}

We change the topology of the charger coils from two coils to four coils in a two-by-two structure as shown in Fig. \ref{fig:4coils}. The surface area covered by the coils remains the same as well as the overlap between adjacent coils to minimize coupling between them.
As a result, the E-field is reduced to \SI{1032}{V/m}, as shown in Fig. \ref{fig:E4coils}. However, it needs to be noted that the H-field does have a different distribution now and that the magnitude is slightly lower at \SI{5.88}{A/m}. A \newacronym{tedz}{TEDZ}{transmission efficiency dead-zone}\acrfull{tedz} can be seen, in Fig. \ref{fig:H4coils} exactly in the middle of the charger, where the four coils overlap. This is as expected seeing that the fields of the four separate coils cancel each other out. Depending on the application this can be an issue to keep in mind. A solution can be, if the location of the implant is known, to only excite two coils at a time as mentioned in Section \ref{sec:coiltop} or change the phase of the magnetic fields of the coils. This last approach is demonstrated in Fig. \ref{fig:phase1} where a phase difference of \SI{90}{}° is introduced in the magnetic field of coil 1 compared to the other transmit coils 
As a result, there is no \acrshort{tedz} visible in the middle of the charger anymore. The magnitude of the H-field is slightly higher, \SI{6.46}{A/m} and the E-field is affected as well, it is if further reduced to \SI{929}{V/m}, as shown in Fig. \ref{fig:phase1E}. If coils 1 and 3 both have a phase difference of \SI{90}{}° compared to coils 2 and 4 (coils 1 and 3 have thus the same phase), the field looks different again, more like the two coil prototype, see Fig. \ref{fig:phase12H}. This results in an even higher H-field of \SI{7.98}{A/m} and an E-field of \SI{989}{V/m}, see Fig. \ref{fig:phase12E}. If the location of the implant is thus known, one can adjust the phase difference not only to move the \acrshort{tedz} away from the implant location and reach an overall higher H-field magnitude and to minimize the E-field magnitude.

The next step would be to investigate whether the E-field can be reduced even more when we go from a two-by-two coil topology to a three by two for instance and so on. As a result, the tuning will again get more and more difficult and the way to address the coils is an extra complexity as well. Either this requires more drivers, one for every coil, or a switching circuit to resolve this.
We therefor first look into combining the three previous techniques.

\subsection{Combination of Techniques}
\begin{figure}[htbp]
\centerline{\includegraphics[width=\linewidth]{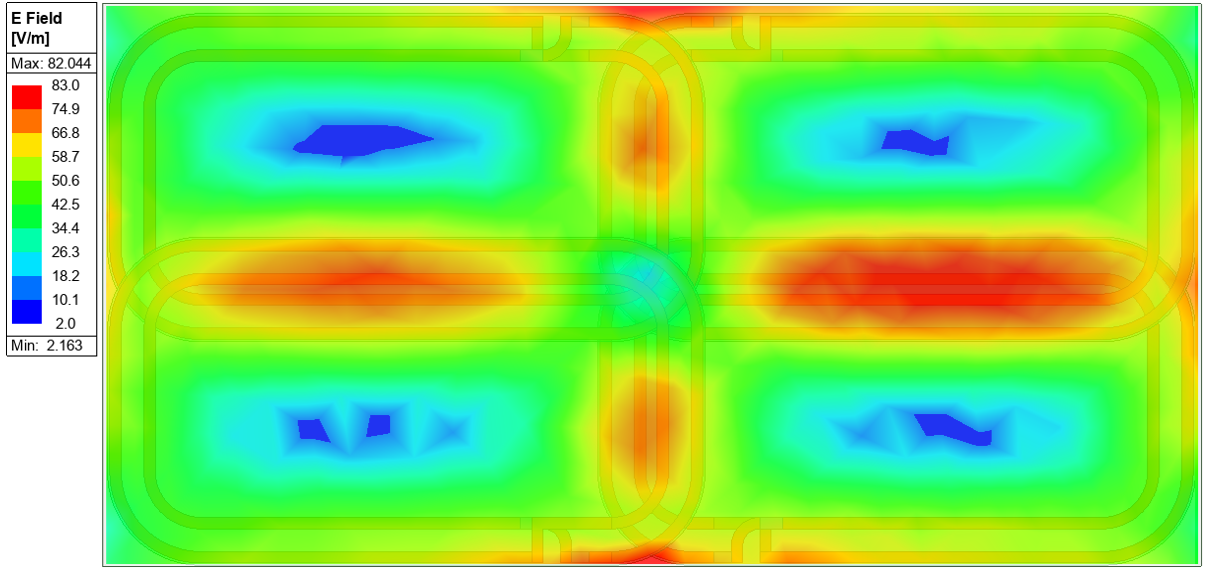}}
\caption{E-field of the charger prototype with the three techniques combined.}
\label{fig:final}
\end{figure}

If we combine the three techniques discussed earlier in this section we get a charger prototype existing of a two-by-two grid of two turn coils with each sixteen tuning capacitors distributed over the traces of the coils and a \SI{1}{mm} $BaTiO_3$ shield \SI{1}{cm} above the charger. As shown in Fig. \ref{fig:final} 
, the E-field of this prototype is reduced to \SI{82}{V/m}, just under the limit. The H-field still reaches \SI{8}{A/m}, we thus adhere to the Canadian RSS-102 limit of \SI{83}{V/m}. The electric field is within limits, the magnetic field is unaltered but what has been overlooked is the overall \acrlong{pte}. To evaluate this aspect, the power consumption of the new prototype needs to be checked. The current that is simulated through the transmit coils in this configuration is \SI{1.88}{\arms} which is indeed higher than the initial \SI{1.24}{\arms}. With the applied voltage of \SI{10}{V} at the lumped port this results in an initial power consumption of \SI{12.4}{W} and a new power consumption of the latest simulation of \SI{18.8}{W}. The overall \acrshort{pte} decreased thus slightly.

\section{Results and Discussion} \label{discussion}
Initial simulations show that the mitigation techniques discussed in this paper all have a significant impact on the magnitude of the electric field emitted by a comfortable wireless charger for medical implants. However, the Canadian RSS-102 standard limit of \SI{83}{V/m} is only met if all three techniques are combined.

The shielding technique has the ability to reduce the E-field by 65 \% if the right material is available. A layer of 1 mm thick with a $\epsilon_r$ of 900, $\tan \delta_e$ of 0.01, $\mu_r$ of 1 and $\tan \delta_m$ of 0.0001 can reduce the E-field from the initial \SI{1416}{V/m} to only \SI{496}{V/m} without altering the magnetic field. The $\tan \delta_e$ has less of influence than the $\epsilon_r$ on the E-field and is thus less critical to take into account when choosing a material. The same goes for the $\tan \delta_m$, as long as it is rather small ($\approx 10^{-4}$) and $\mu_r$ is around 1, the magnetic field will remain the same as without the material. The material that best corresponds to those values is $BaTiO_3$. However, this is a rather expensive material and it is acquired in the form of crystals, which may be an issue to make it into a layer to go on top of the charger prototype. With the use of Fig. \ref{gr:perm} different materials can be sought depending on the application and the desired field reduction.

The distributed capacitors technique is very effective at reducing the electric field, however, it is a more tedious technique as the tuning of the system becomes more cumbersome because of the multiple capacitors. From Fig. \ref{gr:C} we see that doubling the amount of capacitors reduces the E-field with 30 to 50\% each time. In the application of this research, we reached an electric field as low as \SI{231}{V/m} with sixteen capacitors distributed on the turns of the transmit coils. We did not add any more capacitors seeing that the effort of tuning the system became too cumbersome for the small field reduction that we might still get. Instead, we looked at the third and last technique of changing the prototype typology from two coils to four coils in a two-by-two structure. This resulted again in a reduction of around 30\%. In the future it would be interesting to study different topologies such as six coils in a two-by-three structure but just as was the case with the previous technique, the tuning of the system became too cumbersome as the coils were coupled to one another. Additionally driving the charger system becomes more complex as well. Either more drivers are needed, one for every coil, which takes up more space in the charger system and becomes more expensive. Either the design of a switching circuit is required to be able to drive all coils with for instance only two drivers.
Therefor, we opted to look at the combination of the best outcomes of all three techniques. The prototype is now a two-by-two coil charger with sixteen capacitors divided on the turns of each coil and a layer of $BaTiO_3$ on top of it. The electric field is now reduced to \SI{82}{V/m}, just below the \SI{83}{V/m} limit of the Canadian RSS-102 standard. The magnetic field remained unchanged, still reaching \SI{8}{A/m} at a height of \SI{10}{cm} above the charger. Table \ref{tab:E} summarizes the effect of different techniques on the E- and H-field.

\begin{table}[!h]
    \caption{Summary of the field strengths and the challenges of the different prototypes.}
    \label{tab:E}
    \centering
    \renewcommand{\arraystretch}{1.2}
    \begin{tabular}{@{}lrrr@{}}
    \toprule
    Technique & E & H & Challenge \\
    & [V/m] & [A/m] & \\
    \hline
    Initial prototype & 1416 & 8.48 & \\
    Initial prototype with BaTiO\textsubscript{3} & 496 & 8.35 & Crystals \\
    2 coils, 2 C & 842 & 8.44 & \\
    2 coils, 4 C & 519 & 8.32 & \\
    2 coils, 8 C & 376 & 8.30 & \\
    2 coils, 16 C & 231 & 8.57 & Tuning \\
    4 coils & 1032 & 5.88 & Coupling \\
    4 coils, phase difference coil 1 90° & 929 & 6.46 & \\
    4 coils, phase difference coil 1 and 3 90° & 989 & 7.98 & \\
    4 coils, 8 C & 363 & 7.19 & \\
    4 coils, 16 C & 250 & 6.61 & \\
    4 coils, 16 C, BaTiO\textsubscript{3} & 82 & 8.12 & \\
    \bottomrule
    \end{tabular}
\end{table}

\section{Conclusion and Future Work} \label{conclusion}
This research identified and quantified effective techniques to mitigate electric field emissions in an inductive wireless charger for medical implants, ensuring compliance with stringent safety standards without compromising on the magnetic field strength and thus power transfer efficiency. Initial simulations of a two-coil transmitter operating at \SI{6.78}{MHz} delivering \SI{8}{A/m} at a distance of \SI{10}{cm}, revealed an unacceptably high peak E-field of \SI{1416}{V/m} at 2 cm from the transmitter, far exceeding the \SI{83}{V/m} regulatory limit by the Canadian RSS-102 standard. This paper investigated three mitigation strategies in detail and found that each contributes significantly to reducing E-field levels without impairing magnetic field strength. In the future it would be interesting to study the effects of these techniques on E-fields in general, without the constraints of the application. More parameters can be considered such as the thickness of the layer of shielding material, the height of this layer above the transmit coils etc.

\ifCLASSOPTIONcaptionsoff
  \newpage
\fi

\end{document}